\documentclass[aps,prx,twocolumn,floatfix,
superscriptaddress]{revtex4}

\usepackage[T1]{fontenc}
\usepackage[english]{babel}
\usepackage{ae}
\usepackage{wrapfig}
\usepackage{float}
\usepackage{amsmath}
\usepackage{amssymb}
\usepackage{amsfonts}
\usepackage[pdftex]{graphicx}	
\usepackage{array}	
\usepackage{bm}		
\usepackage{dsfont}	
\usepackage{ulem}
\usepackage{color}
\usepackage{hyperref}
\usepackage{units}
\usepackage{gensymb}
\usepackage{ulem}

\renewcommand{\emph}{\textit}

\begin{document} 





\title{Unconventional pairing in single FeSe layers}

\author{Jasmin Jandke,$^{1}$ Fang Yang,$^{1}$ Patrik Hlobil,$^{2,5}$ Tobias Engelhardt,$^{1}$ Dominik Rau,$^{3}$ \\ Khalil Zakeri,$^{3}$  Chunlei Gao,$^{4}$  J\"org Schmalian,$^{2,5}$ Wulf Wulfhekel,$^{1,4}$\\
$^{1}$Physikalisches Institut, Karlsruhe Institute of Technology,
76131 Karlsruhe, Germany\\
$^{2}$Institut f\"ur Theorie der Kondensierten Materie, Karlsruhe Institute of Technology,
76131 Karlsruhe, Germany\\
$^{3}$Heisenberg Spin-dynamics Group, Physikalisches Institut, Karlsruhe Institute of Technology,
76131 Karlsruhe, Germany\\
$^{4}$State Key Laboratory of Surface Physics and Department of Physics, Fudan University,
Shanghai 200433, China\\
$^{5}$Institut f\"ur Festk\"orperphysik, Karlsruhe Institute of Technology,
76344 Karlsruhe, Germany
}
\date{\today}

\begin{abstract}
The pairing mechanism in iron-based superconductors is believed to be unconventional, i.e. not phonon-mediated. The achieved transition temperatures $T_{\rm c}$ in these superconductors are still significantly below those of some of the cuprates, with the exception of single layer FeSe films on SrTiO$_3$ showing a $T_{\rm c}$ between 60 and  \unit[100]{K}, i.e. an order of magnitude larger than in bulk FeSe.
This enormous increase of $T_{\rm c}$ demonstrates the potential of interface engineering for superconductivity, yet
the underlying mechanism of Cooper pairing is not understood. Both conventional and unconventional mechanisms have been discussed.
Here we report a direct
measurement of the electron-boson coupling function in FeSe on
SrTiO$_3$ using inelastic electron scattering which shows that the excitation spectrum becomes fully gapped below $T_{\rm c}$ strongly supporting a predominantly electronic pairing mechanism.
We also find evidence for strong electron-phonon coupling of low energy electrons, which is however limited to regions near structural domain boundaries.
\end{abstract}

\maketitle

\section{Introduction}

Achieving high-$T_{{\rm c}}$ superconductivity through surface and
interface engineering is among the most fascinating recent developments
in the field of complex materials design~\cite{Reyren07,Ueno11,Mizukami11,Ye12,Saito16}. 
FeSe is the simplest Fe-based high-$T_{{\rm c}}$ superconductor showing superconductivity
in bulk below ${\approx \rm 8} \, K$ at ambient pressure~\cite{Hsu08}.
It consists of two-dimensional Fe$_{2}$Se$_{2}$ layers, weakly bound
by van-der-Waals forces. 
Most notably, the prominent increase of $T_{{\rm c}}$ to the highest
transition temperatures (between 60 and \unit[100]{K}) within the family of the iron-based systems  
have been reached in a single Fe$_2$Se$_2$ layer on Nb-doped
SrTiO$_{3}$~\cite{Wang12,Liu12,He13,Tan13,Lee14,Ge15}. The success of this materials-design strategy intimately relies on a clear understanding of the
key physical properties, changed through the interface. 
It has been
established for single-layer Fe$_2$Se$_2$ that the substrate suppresses nematic
order, changes the lattice constant, modifies the electronic structure,
and supplies additional charge carriers (for a review see for example Ref. \cite{Liu15,Huang17}).

\begin{figure}[htb]
\centering
\includegraphics[width=0.8\columnwidth]{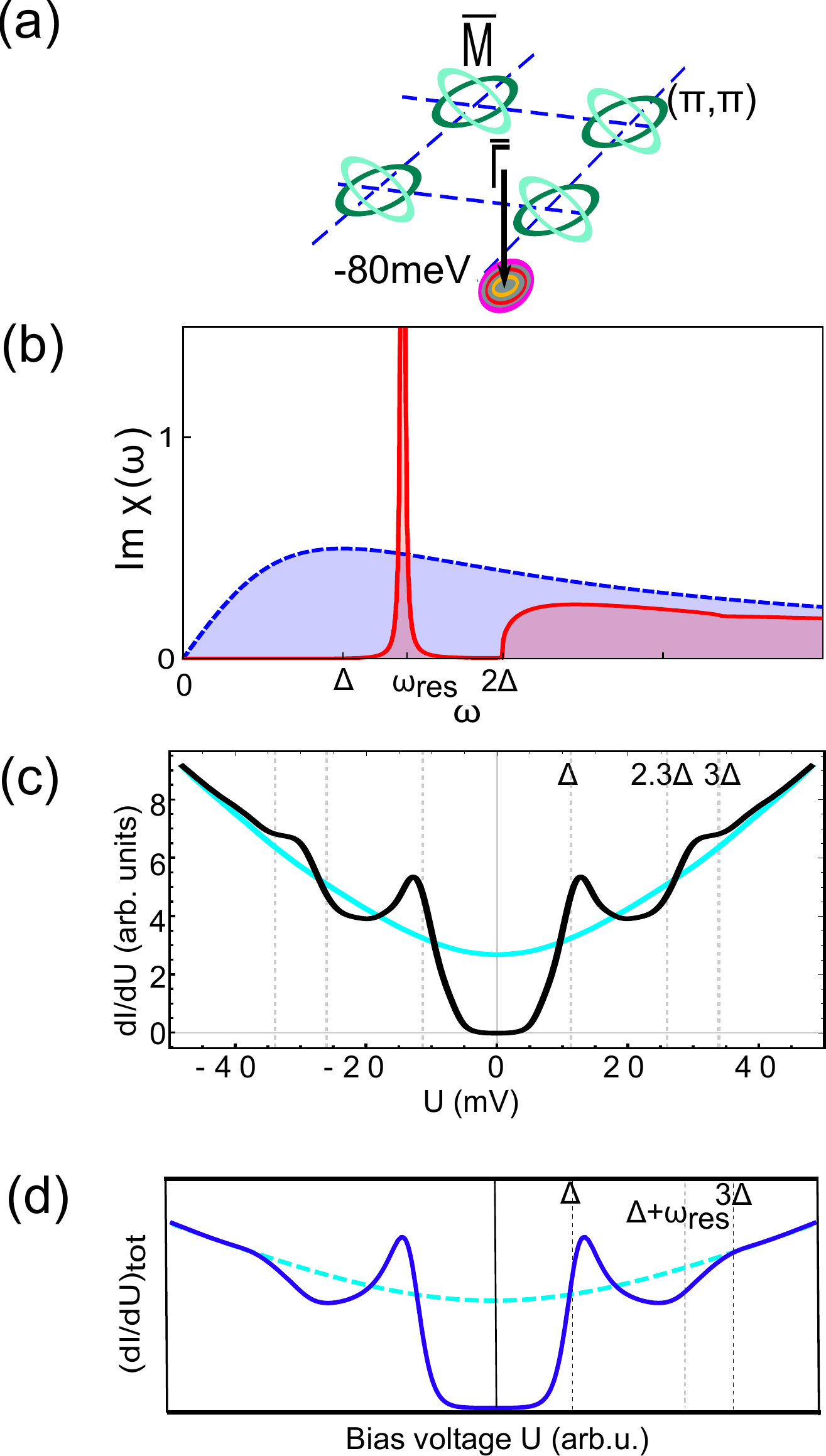} 
\caption{ (a) Sketch of the band structure of a single layer FeSe on SrTiO$_{3}$  with two electron bands around the zone
corner and three hole bands in the zone center located \unit[80]{meV} below the Fermi energy. 
(b) Sketch of the magnetic excitation spectrum ${\rm Im}\chi(\omega) $ above (blue dotted curve) and below (solid red curve) the transition temperature $T_{\rm c}$ \cite{Hlobil17}. 
(c) Symmetrized experimental $dI/dU$ spectra in the superconducting/normal state (black/blue) recorded at 0.9/62 K. The positions $\Delta,2.3\Delta,3\Delta$
are shown by gray dashed lines. The spectra are normalized
to the same differential conductance at 50 meV. (d) Theoretical differential conductivity between a normal conducting tip and a spin-fluctuation driven superconducting sample above (dashed blue curve) and below (solid blue curve) $T_{\rm c}$. $\omega_{\rm res}$  is the energy of the resonance mode. 
}
\label{F1} 
\end{figure}

First principles calculations of the transition temperature of bulk FeSe solely
based on conventional electron-phonon coupling predict a much lower
critical temperature than the observed one~\cite{Li14}.  In addition,  neutron scattering experiments below $T_{{\rm c}}$ revealed a magnetic resonance mode within the superconducting gap \cite{Wang2016}.  The resonance mode is a consequence of the superconducting coherence factors in the BCS-type wave function in a state with sign changing order parameter \cite{Abanov1999,Eschrig2000,Abanov2002,Eschrig2006,Korshunov2008}.
Such a change in sign of the pairing wave function reveals an unconventional pairing state and is a strong evidence for an electronic pairing mechanism \cite{Abanov1999,Eschrig2000,Abanov2002,Eschrig2006,Korshunov2008,Wang2016}. In a number of strongly coupled superconductors, the ratio of the resonance mode and the superconducting gap was found to be  close to the universal value $\omega_{\rm{res}}\sim 1.3 \Delta$ \cite{Yu09}, a behavior owed to the strong coupling nature of this excitonic bound state \cite{Hlobil2013}.
In many iron-based materials, the change of sign in the gap of s$_{\pm}$-symmetry is believed to occur between the electron and hole pockets,
and pairing be mediated by spin fluctuations near the antiferromagnetic ordering vector of stripe type ordering \cite{Mazin08}.  Since stripe type magnetic fluctuations have been observed in the low-$T$ nematic phase~\cite{Wang2016},  bulk FeSe is likely an unconventional superconductor with similar mechanism. However, in the single-layer FeSe samples, the  electronic structure is significantly changed from the bulk material with hole pockets pushed down to about \unit[80]{meV} below the Fermi level. Thus, the Fermi surface consists of only the two electron bands near the
zone boundary~\cite{Liu12,He13} (see Fig. \ref{F1}a). This remarkable observation challenges the above consideration of the gap symmetry for  the monolayer system, as the hole pockets well below the Fermi energy do not evolve a gap. The observation of shadow bands in angle-resolved photoemission spectroscopy (ARPES) data has been considered as an indication of a conventional pairing by vertically polarized optical phonons near the zone center localized at the TiO$_{2}$ interface layer~\cite{Lee14,Peng14,Xie15,Wang17_b,Zhang17,Rebec17,Wang16}. 

The interesting observation of the gap-variation with oxygen isotope substitution on the one hand~\cite{Song_2017} and estimations for comparatively small electron-substrate-phonon coupling constants on the other~\cite{Nekrasov_2017_b} reveal the complexity of this open problem. Recently an alternative interpretation of shadow bands due to strong coupling between surface, Fuchs-Kliewer, vibrations and the escaping photoelectron has been proposed~\cite{Li_2017}. 

In order to disentangle the impact of electronic  and phononic modes for the mechanism of superconductivity, a natural strategy is to compare their excitation spectrum, weighted by appropriate interaction matrix elements, above and below $T_{\rm c}$. While both are affected by superconductivity, electronic collective modes undergo much more dramatic changes. A superconducting gap $\Delta=\Delta_{{\bf k}_{\rm F}}$ in the electronic spectrum gives rise to a gap  $\left| \Delta_{{\bf k}_{\rm F}}\right|+\left| \Delta_{{\bf k}_{\rm F}+{\bf q}}\right| \approx 2\Delta$ of an electronic mode of momentum ${\bf q}$. While broad and overdamped above $T_{\rm c}$, sharp excitonic states inside the gap are again a clear signature for electronic interactions (see Fig. \ref{F1}b). If the characteristic bosonic momentum is finite and in the spin sector, such sharp structures are strong evidence that the sign of  $\Delta_{{\bf k}_{\rm F}}$ and $ \Delta_{{\bf k}_{\rm F}+{\bf q}} $  are distinct. As discussed, these so called resonance modes typically arise in the spectrum of collective electronic excitations at $\omega_{\rm{res}} \approx1.3 \Delta$, as discussed above.
Thus, in order to uncover the role of electronic pairing interactions one has to gain information about the collective mode spectrum of single layer materials above and below $T_{\rm c}$.
We here used two inelastic electron scattering techniques to experimentally determine the coupling between electrons and bosonic excitations. On the one hand, we employed spatially resolved inelastic tunneling spectroscopy (ITS) between the tip of a scanning tunneling microscope (STM) and the sample at high energy resolution. 
On the other hand, we investigated inelastic scattering in momentum space with high resolution electron energy loss spectroscopy (HREELS). Results from the two complementary techniques obtained on the same sample are combined in order to disentangle the role of electronic and phononic excitations in Cooper pairing.

\section{Inelastic tunneling spectroscopy}

In STM, the electron current $I$ tunneling between an atomically sharp tip and the sample is recorded. As has been shown recently in the context of conventional~\cite{Schackert15,Jandke16} and unconventional~\cite{Hlobil17} superconductors, the tunneling conductance between a normal conducting tip and a superconducting material consists besides the well-known elastic contributions $\sigma_{\text{el}}$
of significant inelastic contributions $\sigma_{\text{inel}}$: $dI/dU=\sigma_{\text{tot}}=\sigma_{\text{el}}+\sigma_{\text{inel}}$. Due to the spatial confinement of the electrons in the tip apex, the
wave vector of the tunneling electrons is widely spread. As a consequence,
the elastic contribution to the tunneling conductance is proportional to the electronic density of states (DOS) of the sample, as has been shown by Tersoff and Hamann \cite{Tersoff1985}. Similarly, the
inelastic contribution to the tunneling conductance is given by the integral of the scattering probability of electrons from bosons over both momentum- and energy-space up to the maximally allowed energy loss, i.e. the bias voltage of the junction. 
In the normal conducting state of a superconductor, the 
DOS of the electrons is nearly constant on the energy scale of the bosonic excitations. Thus, the elastic contributions vanish in the second derivative of the tunneling
current $d^{2}I/dU^{2}$ in the normal state and only inelastic contributions enter. 
In this case, the optical theorem links the inelastic scattering processes of a specific energy loss by inelastic creation of bosons to the imaginary part of the forward scattering amplitude for electron-electron scattering mediated by virtual boson exchange.
Thus the measured second derivative of the tunneling current is given by \cite{Hlobil17}:
\begin{equation}
d^{2}I/dU^{2}\propto
{\rm Im}\chi(\omega)  g(\omega)^{2},
\end{equation}where $\chi(\omega)$ is
the response function or excitation spectrum and $g(\omega)$ is the electron-boson coupling matrix element.
The latter term reflects the fact that ITS only detects bosonic excitations that actually couple to the electron liquid at the relevant energy scale. 
For conventional superconductors, $d^{2}I/dU^{2}$ is then proportional to the so called Eliashberg function
$\alpha(\omega)^{2}F(\omega)$, as has been shown e.g. for Pb films \cite{Schackert15}. In this conventional superconductor, $\alpha(\omega)$ is the electron-phonon coupling constant and $F(\omega)$ is the imaginary part of the lattice response function, i.e. the density of states of the phonons.
In case the bosonic spectrum
is predominantly due to electronic excitations, $d^2I/dU^2$ of the normal state corresponds to the overdamped,
collective modes, which quickly decay into particle-hole excitations near the Fermi energy (Fig. \ref{F1}b).
As there is no electronic gap in the normal state, inelastic contributions to $dI/dU$ appear at an onset bias voltage that equals the energy of the bosonic excitations (phonons or electronic excitations). Note, that at finite temperatures, also $dI/dU$ at zero bias contains an inelastic contribution due to thermally excited tunneling processes. Both, the phonon spectrum and the spectrum of potential spin excitations in the normal state are not gapped and the expected $dI/dU$ signal displays a V- or U-shape \cite{Kirtley90}. Therefore, only from ITS experiments in the normal state, phononic or spin excitations cannot be distinguished.

The situation fundamentally changes in the superconducting state. First of all, the electronic spectrum develops a gap $\pm\Delta$ around the Fermi energy and at least a bias voltage of $\pm e \Delta$ needs to be applied to add or remove a single electron from the superconductor by tunneling.  
Inelastic contributions caused by electron-phonon coupling thus lead to signals on top of the
BCS DOS at voltages corresponding to the phonon energy $\hbar \omega_{\rm ph}$ shifted by the gap $\Delta$~\cite{Scalapino66,Jandke16}.
If, however, the bosonic excitations are of electronic nature, additionally the gap in the electronic excitation spectrum needs to be overcome. Thus the generic expectation is that inelastic excitations of electronic nature appear at voltages beyond $\pm 3\Delta$ or for a sign-changing gap function at slightly lower energies $\Delta + \omega_{\rm {res}} \approx 2.3 \Delta$ due to the resonance mode.
This way, phonon- and electronic excitation mediated superconductivity can be distinguished, experimentally.

In Fig.~\ref{F1}c we show experimental $dI/dU$ spectra for single layer FeSe on STO in comparison to the theoretical model of spin fluctuation driven superconductivity \cite{Hlobil17}, shown in Fig.~\ref{F1}d. In this system, the electronic DOS is not a constant but increases with energy, i.e. the tunneling spectra are tilted. In order to remove this band-structure effect, we
take the standard approach and symmetrize all spectra. This suppresses
the energy dependence of the DOS to first order. The differential
conductance in the normal state, i.e. above the transition temperature which in our case is about 55 K,
is shown in blue. The parabolic increase caused by the inelastic contributions
is a characteristic feature of the normal state~\cite{Kirtley90,Hlobil17}.
The superconducting spectrum (black) shows besides the coherence peaks
at $\Delta\approx\pm$10 meV a reduced differential conductance below
that of the normal state in the energy range above the coherence peaks
up to $\approx2.3\Delta$. Assuming purely elastic tunneling in the BCS approach, the DOS in the superconducting state should always be higher than that of the normal state at energies above the quasi particle peaks at $\pm\Delta$. 
Instead, this reduction in differential conductance demonstrates the presence of inelastic contributions and the formation of a gap in
the bosonic excitation spectrum below $T_{\rm c}$. 
At $\approx2.3\Delta$ $dI/dU$ increases abruptly, stays flat
and rises again at $\approx3\Delta$ to approach the value of
the normal state. Thus, the  behavior is fully in agreement with a spin-fluctuation
mediated superconductivity with a gapped bosonic
excitation spectrum below the resonance mode at $\omega_{{\rm res}}\approx1.3\Delta$~\cite{Hlobil17}. 
Thus, our results strongly support a fully developed superconducting gap, an electronic pairing mechanism with a gap in the excitation spectrum in the superconducting state, and a sign changing order parameter leading to a resonance in the bosonic spectrum at $\approx1.3\Delta$.
In contrast, the formation of a gap in the inelastic part of the the conductance clearly speaks against phonon mediated superconductivity, as the phonons would not develop a gap but would instead show a softening.

\begin{figure}
\centering 
\includegraphics[width=1\columnwidth]{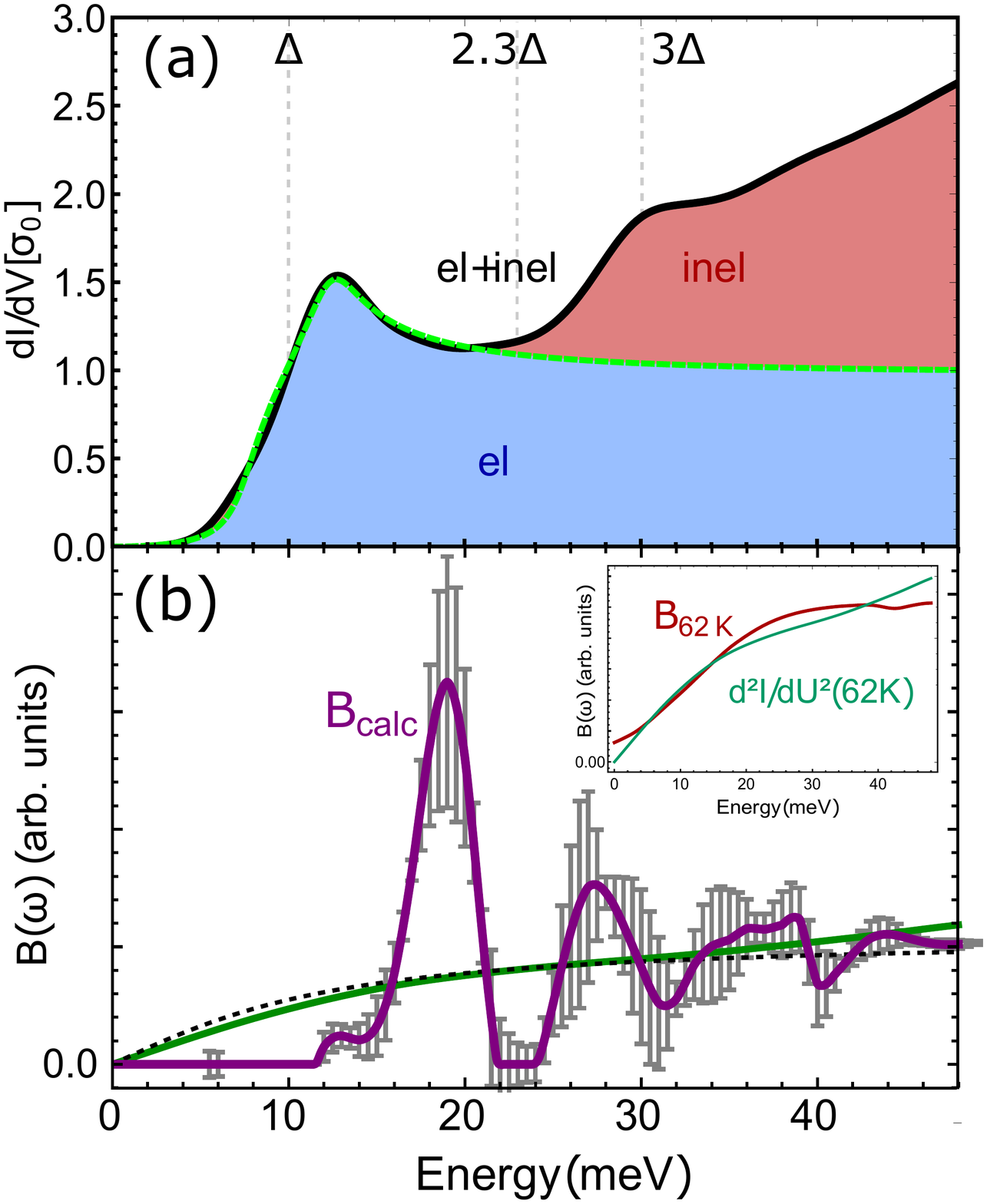}
\caption{(a) Symmetrized tunneling spectrum in the superconducting state recorded at 0.8 K (black curve)
together with a fit of the elastic contribution $\sigma^{\text{el}}$ due to the anisotropic gap function 
(blue area). The red area indicates the inelastic contributions to the conductance. (b) Initial trial function (black dotted curve), experimental inelastic $d^{2}I/dU^{2}$ spectrum in the normal state recorded at \unit[62]{K} (green) and 
reconstructed $B(\omega)$-spectrum in the superconducting state (purple). The gray error bars indicate the uncertainty due to the gap anisotropy.
The inset illustrates that the measured $d^{2}I/dU^{2}$
in the normal state deviates from the thermally broadened $B(\omega)$-spectrum (red). }
\label{Fig2} 
\end{figure}

Now, let us take a step further and obtain from the observed tunneling spectrum
the weighted bosonic spectrum $B(\omega)=g(\omega)^{2}{\rm Im}\chi(\omega)$. In distinction to the Eliashberg function
$\alpha(\omega)^{2}F(\omega)$ for phonon-mediated superconductivity, ${\rm Im}\chi\left(\omega\right)$
for electronic pairing is heavily renormalized when entering the superconducting
state. In order to obtain $g(\omega)^{2}{\rm Im}\chi(\omega)$, we take the symmetrized
spectrum in the superconducting state with its characteristic features
and estimate the elastic conductance by an anisotropic gap fit in the energy range where inelastic tunneling is forbidden, i.e for $|eU|<\Delta+\omega_{{\rm res}}\approx 2.3 \Delta$. On the assumption of the existence of two elliptical electron pockets at the corner near the $\bar{M}$-point that are related by rotation symmetry, we  
fit the gap by a model function 
 $\Delta(\varphi)=\Delta_0+a\cdot\text{cos}(2\varphi)+b\cdot\text{cos}(4\varphi)$  
 resulting in an average gap of $\Delta_0=10.0\pm 0.3$ meV, with anisotropies $a=2.4 \pm 1.1$ meV and $b=0 \pm 2.2$ meV in good agreement with ARPES results of Ref.~\cite{Zhang16_c}.
As shown in Fig. \ref{Fig2}a, the fit (dashed green curve) nicely reproduces the gap and the coherence peaks,
i.e. it is in agreement with elastic tunneling. In a next step,
the experimental inelastic conductance can be estimated from
$\sigma_{\rm inel}^{\rm exp}(eU)=\sigma_{\rm tot}^{\rm exp}(eU)-\sigma_{\rm el}$, as
indicated in red.

In order to extract $B(\omega)$ from $\sigma_{inel}$,
we start with a trial function inserted into the expression
for the inelastic tunneling current given in Ref. \cite{Hlobil17}
for the inelastic contribution to the conductance and generate an
improved expression for $B(\omega)$ taking into account also the anisotropy of $\Delta(k)$, where $k$ denotes the wavevector. 
We use the improved expression to repeat this procedure
until convergence is reached. For details see the supplementary material.
The initial guess trial function is a smooth function and reflects
the overdamped spin fluctuations of the normal state (black dotted
curve in Fig. \ref{Fig2}b). It is chosen to be similar to the measured
inelastic $d^{2}I/dU^{2}$ spectrum in the normal state (green curve in Fig. \ref{Fig2}b), measured above $T_{\rm c}$ (\unit[60]{K})
using $d^{2}I/dU^{2}\propto d\sigma_{\text{inel}}/d(U)\propto g(\omega)^{2} \rm Im \chi(\omega)$,
valid above $T_{\rm c}$. 

The result determined in this way is proportional to ${g(\omega)^{2}\rm Im}\chi(\omega)$
in the superconducting state (purple spectrum in Fig.~\ref{Fig2}c) and clearly
shows a gap followed by a peak slightly above $1.3\Delta$. Here, $\Delta$ is not a sharp value due to its anisotropy in $k$-space and the uncertainty of the fitting. Within the standard deviation $\Delta(k)$ is in the range between 6.2 and 13.8 meV.
At higher energies, a continuous excitation spectrum is seen. All this is in agreement
with unconventional superconductivity mediated by spin-fluctuations and a sign changing order parameter and the changes of the spin-fluctuation spectrum between the normal
(black dotted curve) and the superconducting state (purple curve) agrees well with
the expected renormalization~\cite{Hlobil17}. Note that the indicated error bars shown in gray reflect the uncertainty of the fit of the anisotropic gap function.
In order to check for consistency,
i.e. that the spectrum acquires the gap below $T_{{\rm c}}$, we thermally smeared the low-temperature
spin-fluctuation spectrum  by the respective Fermi-functions of the electrodes (red curve in inset). The thermal smearing at that temperature
is not enough to completely wash-out the overall shape. Most importantly,
it clearly deviates from the spectrum measured at that temperature
(green curve in inset of Fig.~\ref{Fig2}b). Thus, the bosonic
spectrum shows a gap in the superconducting state but not in the normal state, in agreement with an electronic pairing mechanism in our samples.  

As shown in the supplementary material \cite{Supplementary}, we observed variations in
$\Delta$ from sample to sample or within different areas of the sample. This is also well documented by the many different experimental $dI/dU$ spectra obtained on FeSe/STO with STM in the literature~\cite{Wang12,Li_2014_b,Huang17}.
The observed features in the extracted ${\rm Im}\chi\left(\omega\right)$
are present in all tunneling spectra and show a gap that roughly scales
with $\Delta$ in agreement with the spin-fermion model. 

\begin{figure}[htbp]
\centering
\includegraphics[width=1.0\columnwidth]{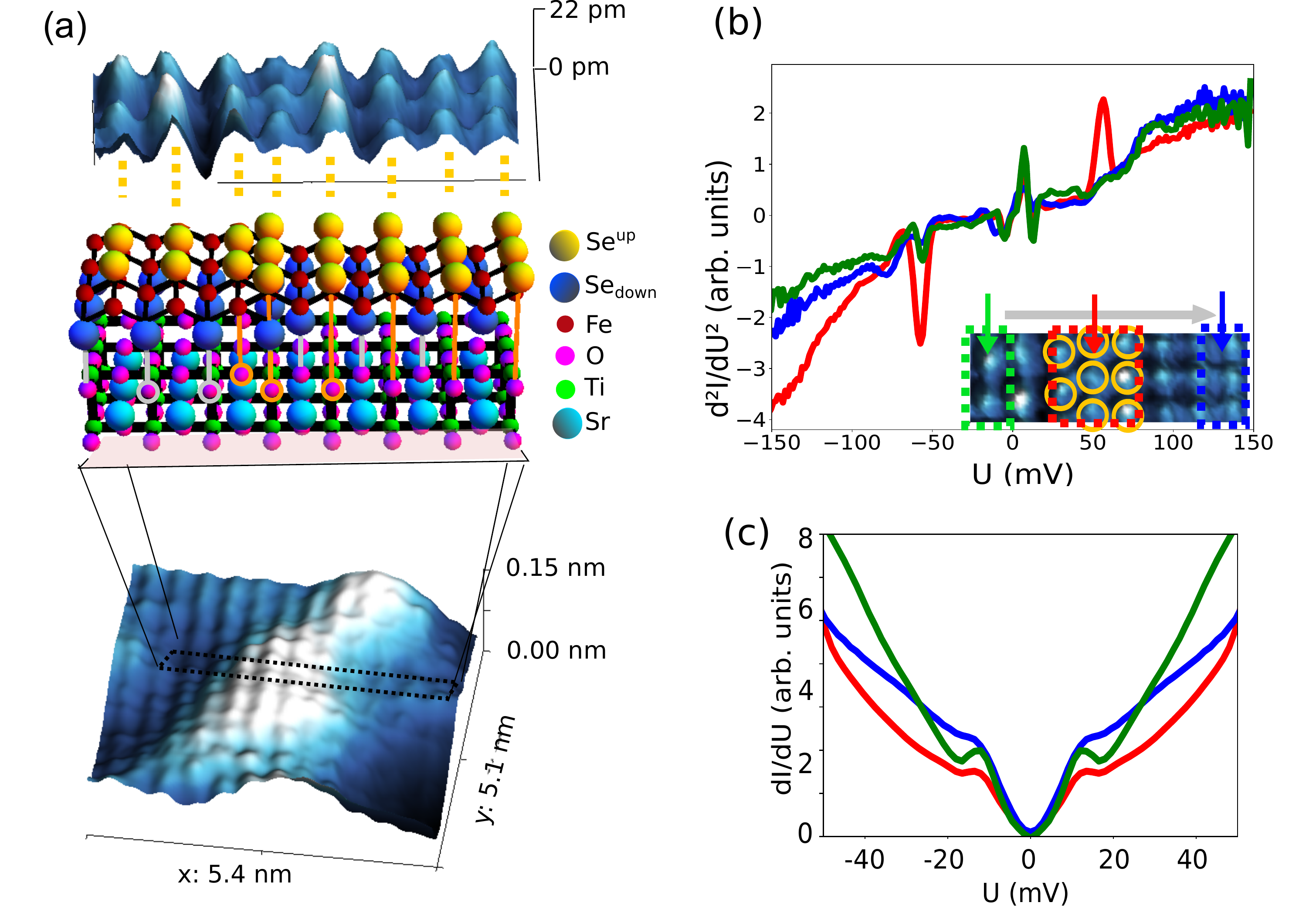}
\caption{(a) STM topography of a dislocation recorded at $U$=\unit[50]{mV} and $T$= 0.8 K. The lower panel shows a magnified image near the boundary of structural domains, shown in the upper panel.  The left and right domains are of different stacking of the FeSe-layer as sketched in the atomic model in the middle panel. This leads to an 1$/$2 unit cell offset along the Fe-Fe direction. (b) The inset shows the same area as the top panel of a) in which the boundary is marked with yellow circles around the detected upper Se atoms. The blue/red/green dashed rectangles mark the areas over which the $d^2I/dU^2$ spectra in (b and c) were averaged. Averaged $d^2I/dU^2$ spectra (I=\unit[2.5]{nA}, U$_{\text{m}}$=\unit[4]{mV}) are shown in solid lines with colors conform to the ones of the inset. (c) Symmetrized $dI/dU$ spectra averaged over the same areas and recorded at the same time as the data in (b).}
\label{fig:phononSTM}
\end{figure}

Similarly, ITS may be used to resolve electron-phonon coupling in the energy range of the STO interfacial phonons.
Interestingly, the ITS data recorded on the FeSe layer are not homogeneous. Fig.~\ref{fig:phononSTM} summarizes these findings. The lower panel of Fig.~\ref{fig:phononSTM}a shows an atomically resolved topography of the FeSe film near a translational domain boundary (white). The domain walls are not straight but meander (see bottom STM topography and the literature \cite{Li_2014_b,Fan15}). Across the domain boundary, the lattice is shifted by half a unit cell, i.e. upper and lower Se atoms are exchanged at the surface as illustrated in the middle panel of Fig.~\ref{fig:phononSTM}a.  Figure~\ref{fig:phononSTM}b shows three different $d^2I/dU^2$ spectra which were averaged over the corresponding marked areas of the inset (top view of upper panel of a). An antisymmetric dip-peak pair is clearly visible around $\pm \unit[60]{meV}$ for the red spectrum, i.e. at the domain boundary. This energy coincides with the energy of the lower optical phonon mode at the the interface  \cite{Zhang16_b,Song_2017,Zhang2018} shifted by $\Delta$. This mode is nearly absent in the areas next to the boundary (blue/green). Furthermore, a significant inelastic excitation around $\pm \unit[90]{meV}$ is missing corresponding to the higher energy optical phonon of the interface \cite{Zhang16_b,Song_2017,Zhang2018}.


This demonstrates that in our samples, the coupling between the low energy electrons and the \unit[50]{meV}-phonon is restricted to the domain boundary area, i.e. is localized, and significant coupling to the \unit[90]{meV}-phonon is observed neither on the domain walls nor within the domains. Most interestingly, the size of the superconducting gap does not change significantly when going through the domain boundary. This is illustrated in Fig.~\ref{fig:phononSTM}d with $dI/dV$ spectra of the gap region where the size of the gap does not vary in agreement with previous studies~\cite{Fan15}.

The fact that electron interaction with the \unit[50]{meV}-phonon is only observed at the boundary, while the superconducting gap size does not change clearly shows that the interfacial electron-phonon coupling of this mode does not boost $T_{\rm c}$ in our sample.  Besides this, ITS measurements showed no indication of strong coupling of low-energy electrons to the \unit[90]{meV} substrate phonons either.

\section{High-resolution electron energy-loss spectroscopy}

HREELS is a powerful technique for the investigation of the surface phonons under UHV conditions.  It allows probing both the excitation energy and the linewidth of all the vibrational modes over the whole surface Brillouin zone.
The monochromatized  low-energy electron beam transfers well-defined energy quanta to the the phonon modes of the system while it  scatters from the surface. This leads to an energy-loss of the electrons after the scattering event. Since the electrons interact with the charge density distribution at the sample surface (and not only with the electron density distribution), the scattering cross section at $T=0$ is directly proportional to the dielectric loss function, ${\rm Im}\{1/\epsilon(\omega , {\bf q})\}$. This quantity is, in turn,  proportional to the imaginary part of the frequency and momentum dependent charge density response function at the surface ${\rm Im}\chi(\omega , {\bf q})$ \cite{Ibach_Mills2013}. In practice the measured HREELS spectra are dominated by the so-called (quasi-)elastic peak at zero energy-loss. However, the inelastic part of the spectra contains all the necessary information regarding the collective excitations.

\begin{figure*}
\centering \includegraphics[width=0.9\textwidth]{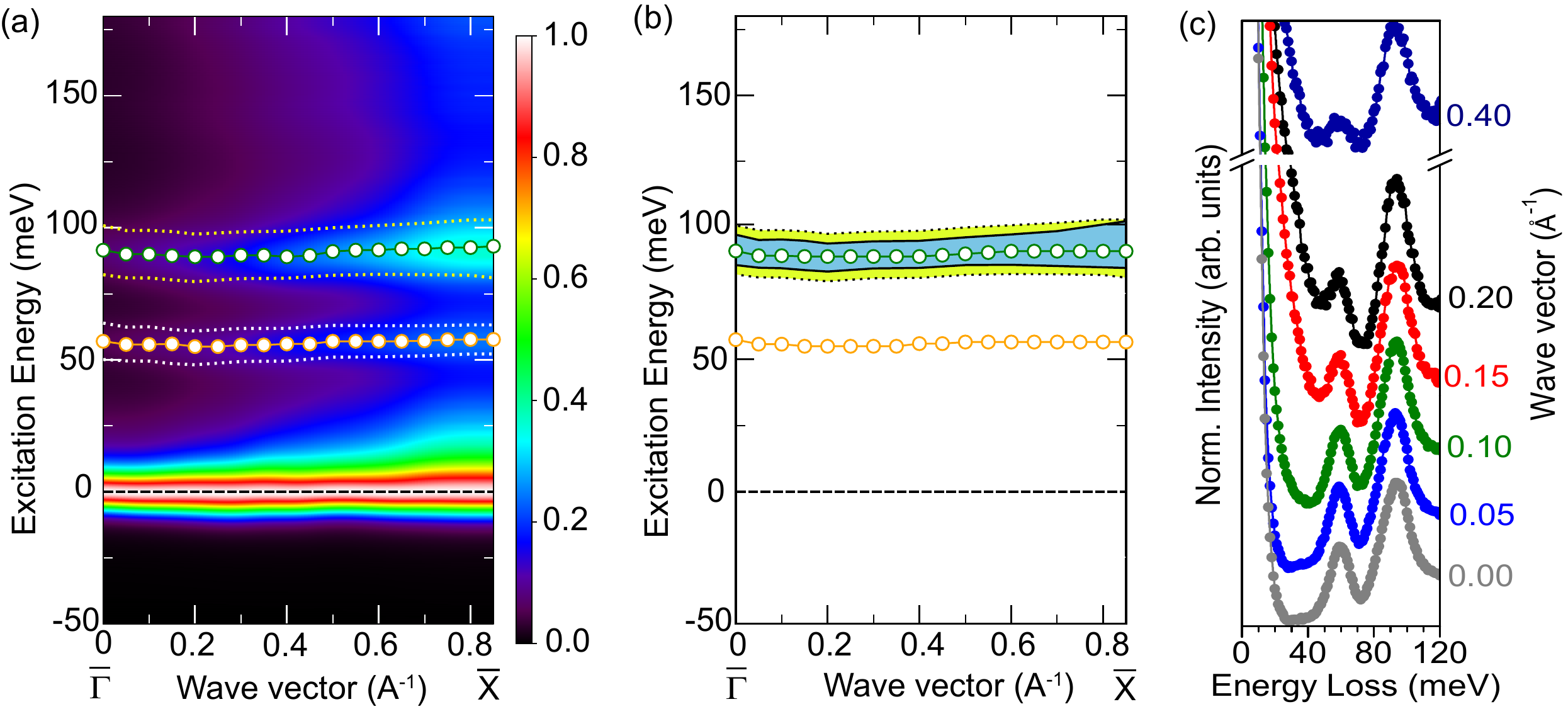}
\caption{(a) Dispersion relation of the interfacial phonons of the FeSe monolayer
on STO recorded at $T=$135 K. The color bar indicates the intensity of the EELS spectra.
All the spectra are normalized to the quasi-elastic peak. The two
phonon branches are indicated by the orange and green circles. 
Dotted lines indicate the full width at half maximum of the phonon
peaks, which includes both the intrinsic linewidth as well as the
experimental broadening. (b) Details of the linewidth of the higher
energy phonon mode versus wave vector. The full width at half maximum corresponds to the shaded area within the dotted lines and the intrinsic linewidth
is represented within the area of the solid black lines. (c) Momentum resolved HREELS spectra near the $\bar{\Gamma}$-point. The spectra are recorded at $T=13$ K and with a HWHM of about 2.8 meV.}
\label{speels}
\end{figure*}

The sample was transferred via a UHV vacuum suitcase from the STM
to the HREELS set-up. A spin-polarized electron beam was used as the
primary beam for this experiment. As phonons are spin-independent
excitations, we only analyze the spin-integrated signal. The measurements were performed at different values of energy resolution (half-width at half maximum of HWHM=2.5--5.5 meV). The spectra were recorded at different temperatures above and below $T_{\rm c}$. The incident electron energy was between 4.2 and 7.25 eV and the degree of spin polarization
of incident electrons was (72 $\pm$ 5)\%. The wave vector $q$ of
excitations is determined by the scattering geometry: $q=k_{i}\sin\theta-k_{f}\sin(\theta_{0}-\theta)$,
where $k_{i}$ ($k_{f}$) is the magnitude of the wave vector of the
incident (scattered) electrons, and $\theta$ ($\theta_{0}$) is the
angle between the incident beam and sample normal (the scattered beam).
The momentum resolution of the spectrometer is 
is about 0.03 \r{A}$^{-1}$ in our experiment \cite{Zakeri14}. Different wave vectors were achieved by changing the scattering angles. 


In order to shed light on the importance of the so-called  Fuchs-Kliewer (FK) phonon modes of the STO in superconductivity~\cite{Lee14,Xie15,Peng14,Wang17_b,Zhang17,Rebec17,Wang16,Zhang16_b, Song_2017,Zhang2018} we performed HREELS experiments on the same sample as the STM experiments. A summery of our HREELS spectra recorded at $T=135K$ and at different wave vectors is provided in Fig. \ref{speels}a. The inelastic part of the spectra is dominated by the presence of the interfacial FK modes, located at the energies of  $\unit[57] {meV}$ and $\unit[91] {meV}$, near the zone center ($\overline{\Gamma}$--point). The results are in agreement with earlier HREELS data reported on the same system \cite{Zhang16_b,Song_2017, Zhang2018}. These two modes show a weak dispersion
while increasing the wave vector from $\overline{\Gamma}$
towards the zone boundary $\overline{{\rm {X}}}$-point. The almost flat dispersion relation of these two modes indicate that these phonon modes are of optical nature.

The more interesting phonon mode is the one observed at the energy of about \unit[91]{meV}, as it has been suggested to be responsible for the replica band observed in the ARPES experiments and for boosting $T_{\rm c}$ \cite{Lee14,Peng14,Wang17_b,Zhang17,Rebec17,ZWang16,Wang16}. The electron-phonon coupling shall in principle manifest itself in the linewidth of the phonon mode that it coupled to the electrons. However the analysis of the phonon linewith as indicated in Fig. \ref{speels}b revealed no significant momentum dependence. The spectra shown in Fig. \ref{speels}a are fitted using a convolution of a Gaussian and a Lorentzian
function representing the instrumental and lifetime broadening of the phonon. The total
and intrinsic linewidth as a function of wave vector is shown in Fig.~\ref{speels}b. The intrinsic linewidth (HWHM) is about 5.5 meV at the $\overline{\Gamma}$--point and
remains nearly constant over a large fraction of the Brillouin zone up to $q=0.6$ \AA$^{-1}$.
Then it gradually increases to about $8.8\pm0.5$ meV at the $\overline{{\rm {X}}}$
point. The nearly $q$-independent intrinsic linewidth excludes particularly strong electron-boson coupling near the
zone center ($\overline{\Gamma}$\textendash point). Most of the density functional based calculations have shown that the strength of the electron-phonon coupling should in principle be the highest at the $\bar{\Gamma}$--point (see for example Refs. \cite{Li2014, Wang2016b}). This means that for investigation of such an effect one requires to recorded the momentum resolved spectra in the superconducting phase in the vicinity of the $\bar{\Gamma}$--point. Such spectra are presented in Fig. \ref{speels}c. The data were recorded at a sample temperature of about 13 K, i.e. significantly below $T_{\rm c}$. The HWHM of the elastic peak for these measurements was about 2.8 meV. The intrinsic linewidth  broadening of the \unit[91]{meV} phonon mode  is about 5 meV at 13 K. As it is apparent from the spectra, one does not observe any obvious change of this value while decreasing the momentum from 0.4 \AA$^{-1}$ towards the $\bar{\Gamma}$--point. These data thus exclude any particular strong coupling of this phonon mode to the electronic states of the FeSe film in the superconducting state. Moreover, the intrinsic linewidth of this mode is about an order of magnitude smaller than the energy of the mode, inducing only a rather weak electron-electron
attraction. 

We note that in addition to the FK phonon modes localized at the FeSe/STO interface we also observe the $A_{1g}$, $B_{1g}$ and $A_{2u}$ phonon modes of the FeSe film. They appear at nearly the same energies as those observed on the FeSe(001) surface of the bulk crystal \cite{Zakeri2017}. Discussion of those phonon modes is out of the scope of the present manuscript. Here the key observation is, that there is no considerable momentum dependence of the linewidth of the FK phonon modes below \unit[0.6]{\AA$^{-1}$}, in distinction to the expectation of forward scattering dominated coupling to the substrate phonons.

This observation is in contrast to the discussed mechanism of the occurrence of replica bands requiring an almost exclusive electron-phonon coupling at the $\bar{\Gamma}$--point \cite{Lee14}. We suggest instead that the appearance of the replica bands in ARPES data is most likely due to the interaction of the photoexcited electrons with the dynamic electric field of the FK phonons of STO above the surface. The dynamic electric field caused by the FK phonon modes generates a rather long rage electrostatic potential which then extends in vacuum above the surface. It is rather straightforward to imagine that the photoexcited electron  feels this potential when it leaves to the vacuum and can therefore lose energy without significant momentum transfer along the sample surface. This scenario is in accord also with the fact that the shadow bands only appear for the occupied parts of the unshifted electronic bands.
Photoexcited electrons in ARPES experiments thus may be prone to this interaction with the STO FK phonons above the surface after the emission process. This would be in analogy to the underlying mechanism which leads to the probing of FK phonon modes in the HREELS experiments, in particular in specular geometry (at the $\bar{\Gamma}$--point). 
In this respect, our data and explanation are fully consistent with the proposal put forward in Ref.~\cite{Li_2017}. Our results suggest coupling to the \unit[90]{meV} phonon only to appear with free electrons in HREELS and ARPES experiments, i.e. electrons that are of higher energy and leave the interface to the vacuum but not low-energy electrons. 
 
\section{Conclusion}
 In conclusion, HREELS experiments indicate a coupling of electrons to the FK phonons mainly for free electrons in front of the surface
 while STM reveals no significant coupling to the FK phonons with low energy electrons, except for structural domain boundaries. In these, however, $T_{\rm c}$ is not boosted by the phonons. Instead, STM experiments reveal that the electron-boson coupling spectrum becomes gapped below $T_{\rm c}$ and a resonance mode appears, which speaks for an all electronic pairing mechanism with sign changing order parameter.

\section*{Acknowledgments}
W.W. and J.S. acknowledge funding from the Deutsche Forschungsgemeinschaft (DFG) with grants Wu 394/12-1 and SCHM 1031/7-1, respectively. K.Z. acknowledges funding from the DFG through the Heisenberg Programme ZA 902/3-1. F.Y. and J.J. acknowledge funding by the Alexander-von-Humboldt Foundation and the Karlsruhe House of Young Scientists, respectively. 

\bibliography{literature}







\renewcommand{\figurename}{\textbf{Figure S}}
\renewcommand{\thefigure}{\textbf{\arabic{figure}}}
\setcounter{figure}{0}

\setcounter{section}{0}

\renewcommand{\thesection}{\textbf{\arabic{section}}}


\begin{center}

\vspace*{1cm}

\LARGE{Supplemental Material}

\vspace*{-0.3cm}

\end{center}
\normalsize

%


%
%
%

\section{Topography}
\begin{figure}[htpb]
\centering \includegraphics[width=.5\textwidth]{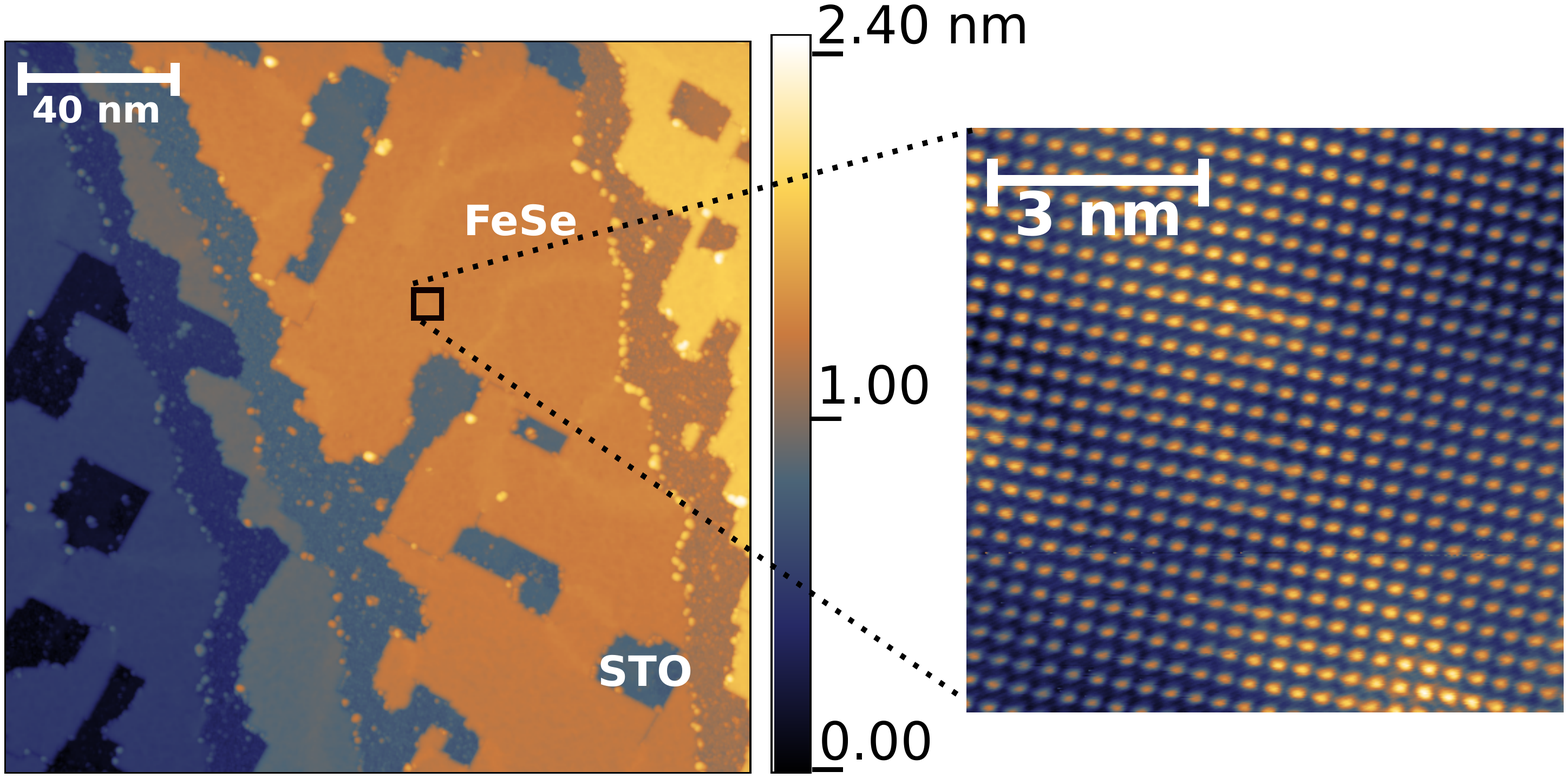}

\caption{STM Topography. Left panel: Overview scan shows STO terraces covered by an almost
complete single layer of FeSe (I=\unit[180]{pA}, U=\unit[1]{V}).
A detailed, atomically resolved topography is presented via the inset
in the right panel. There, the atoms correspond to the upper Se layer
of Se-Fe-Se trilayer. }
\label{S1} 
\end{figure}

A typical topography for an in-situ grown sample is shown in Fig.~S\ref{S1}.
The FeSe-coverage was in this case a bit less than one monolayer.
Within the holes of the film, the bare STO becomes visible.  For the ex-situ samples, a slight degradation of the film quality could
be observed with some persistent impurities arising on the surfaces in agreement with literature ~[1,2].
Nevertheless, for the clean areas, we observed the same spectroscopic
features as for the in-situ sample.

\section{EELS geometry}
\begin{figure}[htpb]
\centering \includegraphics[width=0.45\textwidth]{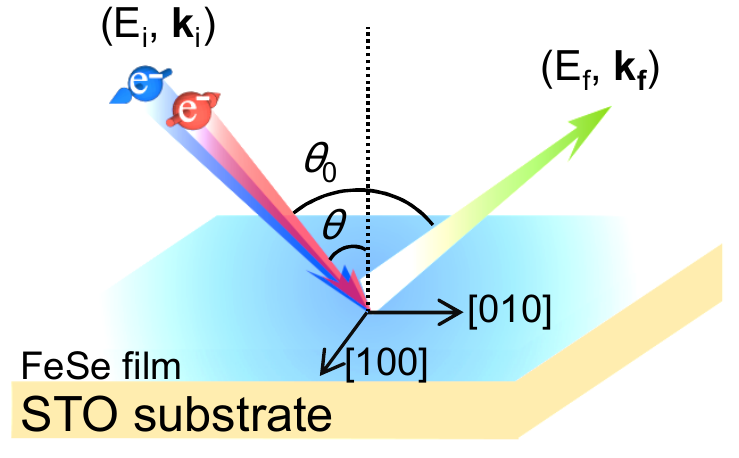} \caption{\textbf{Schematic representation of the scattering geometry in the EELS
experiments.} E$_{i}$ (E$_{f}$) is the energy of the incident (scattered)
beam, $\textbf{k}_{i}$ ($\textbf{k}_{f}$) is the momentum of the
incident (scattered) beam, $\theta$ is the incident angle, and $\theta_{0}$=
80$^{\circ}$ is the angle between the incident beam and the scattered
beam. The scattering plane was along the {[}010{]}-direction. The
phonon wave vector was along the $\overline{\Gamma}$\textendash $\overline{{\rm X}}$
direction of the surface Brillouin zone.}
\label{FigSPEELS} 
\end{figure}

For the electron energy loss spectroscopy measurements schematic representation
of the scattering geometry is given in Fig.~S\ref{FigSPEELS}. 

\section{Electronic structure}
\begin{figure*}[htbp]
\centering \includegraphics[width=0.95\textwidth]{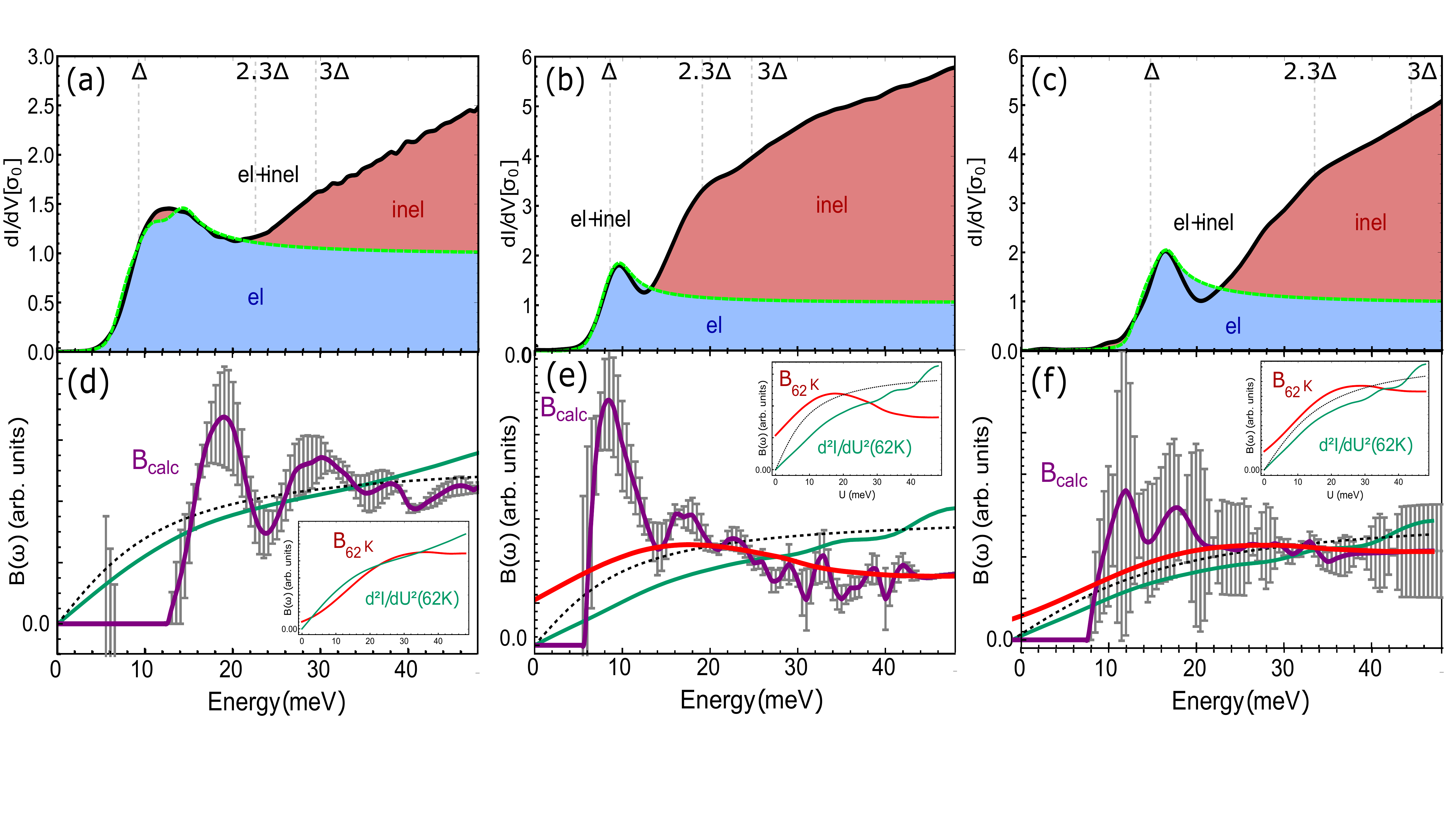} 
\caption{Spectra for different samples and sample-positions. (a),(b),(c): Various measured $dI/dU$ spectra for different samples and sample-positions. Measurement temperatures for curve(a)$/$(b)$/$(c) were  $T=0.8/5.2/\unit[0.8]{K}$ for the superconductig state. Black solid lines represent the measured and symmetrized $dI/dU$ spectra, the gap fits 
($\Delta(\varphi)=\Delta_0+a \cdot\text{cos}(2\varphi)+b\cdot\text{cos}(4\varphi)$) are shown in green ((a): $\Delta_0=\unit[9.73]{meV} \pm \unit[0.48]{meV}$,   $a=\unit[1.95]{meV} \pm \unit[1.26]{meV}$, $b=\unit[2.64]{meV} \pm \unit[1.42]{meV}$, (b): $\Delta_0=\unit[8.35]{meV} \pm \unit[1.26]{meV}$,   $a=\unit[0]{meV} \pm \unit[1.85]{meV}$, $b=\unit[0]{meV} \pm \unit[3.75]{meV}$,  (c): $\Delta_0=\unit[14.7]{meV} \pm \unit[1.05]{meV}$,   $a=\unit[1.70]{meV} \pm \unit[2.58]{meV}$, $b=\unit[0]{meV} \pm \unit[4.23]{meV}$). The extracted $B(\omega)$-spectra are shown in (d), (e), (f) (purple) together with the  starting trial function for the deconvolution
 algorithm (black dotted line) and the measured $d^2I/dU^2$ spectra in the normal state (green) at $T=\unit[62]{K}$. The error bars for the purple curve reflect the uncertainty of the fit of the anisotropic gap function. The calculated temperature broadened $B(\omega)$ spectrum (red) is shown in the insets of (d),(e),(f) and is compared to the  measured $d^2I/dU^2$ spectra in the normal state (green)  $T=\unit[62]{K}$ and the trial function (black dashed). }
\label{S2} 
\end{figure*}

FeSe layers were electronically not completely homogeneous.
The appearance of the superconducting gap varied spatially. Also the
superconducting gap size varied slightly among the samples. Nevertheless, the observed features in the extracted  ${\rm Im}\chi\left(\omega\right)$
are present in all tunneling spectra and show a gap that
roughly scales with $\Delta$ in agreement with the spin-fermion
model. In Figs.~S\ref{S2}a-c, symmetrized $dI/dU$ spectra of three other 
samples and surface positions compared to the spectra shown in the
main text are shown. Different gap sizes are clearly visible. The
blue/red shaded areas illustrate the elastic/inelastic contributions
and the black curve is the total differential conductance.
Figs.~S\ref{S2}d-f represents the corresponding deconvoluted
$B(\omega)$ spectra. A pronounced peak around $\omega_{\text{res}}=1.3\Delta$
is clearly visible in all spectra. Note that as already 
explained in the main text, $\Delta$ is not a sharp value.  It is anisotropic in $k$-space and the performed gap fitting obeys a certain uncertainty. Within the standard deviation $\Delta(k)$ is in the range of $3.64/7.09/9.38$ and $15.82/9.61/\unit[20.02]{meV}$ for the curve in Figs.~S\ref{S2}a$/$b$/$c. For higher energies the
$B(\omega)$ spectra in Figs.~S\ref{S2}d-f  approach a continous background. The insets in  Figs.~S\ref{S2}d-f show B$_{\text{calc}}$ convoluted with the thermal broadening at \unit[60]{K} (red). 
The difference between the measured $d^2I/dU^2$ spectra in the normal state (green) at $T=\unit[62]{K}$ and the trial function is visible.  

Note that experimentally it is not feasible to record pairs of spectra (in the superconducting and the normal state) with exactly
the same tip-sample distance. In order to compare the pairs of spectra,
they have been normalized to the same differential conductance at
50 mV. This assumes that the electronic DOS in both states is identical
for energies far above the gap (elastic tunneling) and identical cross
sections for excitations of high energy bosons by hot electrons (inelastic
tunneling), but neglects thermal assisted inelastic tunneling contributions.
Thus, the normalized spectrum in the normal state is only a lower
bound to the differential conductance recorded for same tip-sample distance.

\section{Theoretical methods for deconvolution of the integrated bosonic tunneling spectrum}
Next we summarize the main steps for the deconvolution of the tunneling
spectra that leads to the boson spectrum $\propto{\rm Im}\chi\left(\omega\right)$.
As shown in Ref.~[3] the inelastic contribution $\sigma_{{\rm inel}}\left(V\right)$ to the
conductivity for $T\ll\Delta$ is given as 
\begin{equation}
\sigma_{{\rm inel}}\left(V\right)=A\int_{-\infty}^{0}d\epsilon\rho_{{\rm sc}}\left(\epsilon\right){\rm Im}\chi\left(\epsilon+eV\right),
\end{equation}
with constant $A=g^{2}\sigma_{0}/\left(D^{2}\rho_{F}\right)$. $\sigma_{0}$
is the normal state conductance of the tip to the substrate. $\rho_{F}$
is the normal density of state and $D$ a characteristic bandwidth of the system.
$\rho_{{\rm sc}}\left(\epsilon\right)$
is the density of states of the superconducting state. In our analysis
we used the finite temperature version of $\sigma_{{\rm inel}}\left(V\right)$
given in Eq.(2) of Ref.~[3].

We determine $\rho_{{\rm sc}}\left(\epsilon\right)$ from a BCS-fit
with angular dependent gap, see main text. $\rho_{{\rm sc}}\left(\epsilon\right)$
also determines the elastic contribution and allows us to subtract
the latter from the total, experimentally determined inelastic conductance
$\sigma_{{\rm inel}}^{{\rm exp}}\left(V\right)$. Starting from an
initial trial for ${\rm Im}\chi\left(\omega\right)$, where we use
a structureless overdamped spectrum realistic to the normal state,
we obtain $\sigma_{{\rm inel}}\left(V\right)$. The difference $\Delta\sigma_{{\rm inel}}=\sigma_{{\rm inel}}^{{\rm exp}}\left(V\right)-\sigma_{{\rm inel}}\left(V\right)$
can now be used to yield a corrected bosonic spectrum 
\begin{equation}
\Delta{\rm Im}\chi\left(\omega\right)=\int d\omega\frac{\delta{\rm Im}\chi\left(\omega\right)}{\delta\sigma_{{\rm inel}}\left(\omega\right)}\Delta\sigma_{{\rm inel}}\left(\omega\right)
\end{equation}
until convergence is reached. The variational derivative is given
as 
\begin{equation}
\frac{\delta{\rm Im}\chi\left(\omega\right)}{\delta\sigma_{{\rm inel}}\left(\omega\right)}=A\int d\epsilon\rho_{{\rm sc}}\left(\epsilon\right)L\left(\epsilon,\omega\right)
\end{equation}
with 
\begin{eqnarray*}
L\left(\epsilon,\omega\right) & = & n_{F}'\left(\epsilon-\omega-eV\right)n_{B}\left(\omega\right)\left(1-n_{F}\left(\epsilon\right)\right)\\
 & + & n_{F}\left(\epsilon\right)\left(1+n_{B}\left(\omega\right)\right)n'_{F}\left(\epsilon-\omega+eV\right)\\
 & + & n'_{F}\left(\epsilon+\omega+eV\right)\left(1+n_{B}\left(\omega\right)\right)\left(1-n_{F}\left(\epsilon\right)\right)\\
 & + & n_{F}\left(\epsilon\right)n_{B}\left(\omega\right)n'_{F}\left(\epsilon+\omega+eV\right).
\end{eqnarray*}

\noindent Here $n_{F}$ is the Fermi-Dirac and  $n_{B}$ the Bose-Einstein distribution.
\newline


\subsection*{Reference}
\noindent [1] Y.-T. Cui, R. G. Moore, A.-M. Zhang, Y. Tian, J. J. Lee, F. T. Schmitt, W.-H. Zhang, W. Li, M. Yi, Z.-K. Liu, M. Hashimoto, Y. Zhang, D.-H. Lu, T. P. Devereaux, L.-L. Wang, X.-C. Ma, Q.-M. Zhang, Q.-K. Xue, D.-H. Lee, and Z.-X. Shen, Interface Ferroelectric Transition near the Gap-Opening Temperature in a Single-Unit-Cell FeSe Film Grown on Nb-Doped SrTiO$_3$ Substrate, Phys. Rev. Lett. {\bf 114}, 037002 (2015). \\

\noindent [2] D. Huang and J. E. Hoffman, Monolayer FeSe on SrTiO$_3$ Annual Review of Condensed Matter Physics \textbf{8}, 311-336, (2017). \\

\noindent [3] P. Hlobil, J. Jandke, W. Wulfhekel, and J. Schmalian, Tracing the electronic pairing glue in unconventional superconductors via inelastic Scanning Tunneling Spectroscopy, Phys. Rev. Lett. \textbf{118}, 167001, (2017).

\end{document}